# New MAX Phase Compound Mo$_2$TiAlC$_2$: First-principles Study

## M. S. Ali[1], M. A. Rayhan[2], M. A. Ali[3], R. Parvin[2] and A. K. M. A. Islam[4,*]

[1]Department of Physics, Pabna University of Science and Technology, Pabna 6600, Bangladesh
[2]Department of Physics, Rajshahi University, Rajshahi-6205, Bangladesh
[3]Department of Physics, Chittagong University of Engineering and Technology, Chittagong-4349, Bangladesh
[4]International Islamic University Chittagong, 154/A College Road, Chittagong, Bangladesh

## Abstract

A theoretical study of the Mo$_2$TiAlC$_2$ compound belonging to the MAX phases has been performed by using the first-principles pseudopotential plane-wave method within the generalized gradient approximation (GGA). We have calculated the structural, elastic, electronic and optical properties of Mo$_2$TiAlC$_2$. To confirm mechanical stability, the elastic constants $C_{ij}$ are calculated. Other elastic parameters such as bulk modulus, shear modulus, compressibility, Young modulus, anisotropic factor, Pugh ratio, Poisson's ratio are also calculated. The energy band structure and density of states are calculated and analyzed. The results show that the electrical conductivity is metallic with a high density of states at the Fermi level in which Mo $4d$ states dominate. Furthermore, the optical properties such as dielectric function, refractive index, photoconductivity, absorption coefficients, loss function and reflectivity are also calculated. Its reflectance spectrum shows that it has the potential to be used as a promising shielding material to avoid solar heating.

## 1. Introduction

Very recently a new ordered ternary carbide Mo$_2$TiAlC$_2$ compound has been identified [1].This layered ternary carbide belongs to the MAX phase which contains Mo. Ti atoms are sandwiched between two Mo-layers that in turn are adjacent to the Al planes resulting in a Mo–Ti–Mo–Al–Mo–Ti–Mo stacking order. The C atoms are in between the Mo and Ti layers [1]. Only other Mo-containing Mo$_2$GaC was known to us [2]. Now a new Mo-containing compound is added to the MAX phases. Before this Liu *et al* [3] synthesized Cr$_2$TiAlC$_2$by sandwiching Cr$_2$AlC and TiC. The crystal structure and chemical composition of the new compound Mo$_2$TiAlC$_2$have been determined by a combination of XRD and HRSTEM measurements [1]. Up to date over 68 MAX phase compounds have been known to us as solid solutions and it is increasing day by day. Some examples are (Nb$_{0.8}$, Zr$_{0.2}$)$_2$AlC [4], Ti$_3$Al$_{1-x}$Si$_x$C$_2$ (0 < x <1) [5], and (Nb$_{0.6}$, Zr$_{0.4}$)$_2$AlC [6], and Mn in (Cr$_{0.7}$, Mn$_{0.3}$)$_2$GaC [7], (Nb$_{0.5}$Ti$_{0.5}$)$_5$AlC$_4$ [8], (V$_{0.5}$Cr$_{0.5}$)$_5$Al$_2$C$_3$ [9] and so on.

The MAX phase compounds show both metallic and ceramics properties. These MAX phases are thermally conductive, easily machinable [10, 11] resistant to thermal shock- they show plasticity at high temperature, highly damage tolerant [12], lightweight, showing no fatigue at high temperature, and oxidation-resistant [13-15]. As a result MAX phase compounds have attracted much attention recently and more and more phases are being synthesized. The general formula of such MAX phase compounds is M$_{n+1}$AX$_n$, (MAX) where n = 1, 2, 3 etc., M is an early transition metal, A is a group XIII - XVI element and X is carbon and/or nitrogen [16-18].

To the best of our knowledge the elastic, electronic, and optical properties of the new $Mo_2TiAlC_2$ layered structure are still unknown. Optical properties (dielectric function, refractive index, extinction coefficient, absorption coefficient, energy loss function, reflectivity, and photoconductivity) in light phenomena and these properties are very important in modern technological system. Therefore, we are interested and encouraged to investigate the different properties of the new $Mo_2TiAlC_2$ compound in this paper.

## 2. Computational Methods

All calculations have been carried out using CASTEP code based on the density functional theory (DFT) [19-22]. Furthermore, the ultrasoft pseudopotential formalism of Vanderbilt [23] is used to simulate the interactions of valence electrons with ion cores, and the electron wave function is expanded in plane waves up to an energy cutoff of 500 eV for all calculations. The exchange–correlation energy is evaluated using the GGA of the Perdew–Burke–Ernzerhof for solids (PBEsol) formalism [22], which is dependent on both the electron density and its gradient at each space point. To search the ground state, a quasi-Newton (variable-metric) minimization method using the Broyden–Fletcher–Goldfarb–Shanno (BFGS) update scheme [24] is utilized, which provides a very efficient and robust way to explore the optimizing crystal structure with a minimum energy. For the sampling of the Brillouin zone [25] a 13×13×2 Monkhorst–Pack mesh is employed [26]. Geometry optimization is achieved using convergence thresholds of $10^{-5}$ eV/atom for the total energy, 0.03 eV/Å for the maximum force, 0.05 GPa for the maximum stress and $10^{-3}$ Å for maximum displacement. By utilizing a set of homogeneous deformations with a finite value under linear proportion, the resulting stress can be calculated with respect to the optimized crystal structure [27]. The elastic coefficients can be finally determined through a linear fit of the calculated stress as a function of strain, where four strain amplitudes are used with the maximum value of 0.3%.

## 3. Result and Discussion

### 3.1 *Structural properties*

$Mo_2TiAlC_2$ crystallizes in the hexagonal structure [1] with space group $P6_3/mmc$ (194). The unit cell has 12 atoms containing two formula units. The fully relaxed structure is obtained by optimizing the geometry with respect to lattice constants and internal atomic positions. The optimized Mo atoms are situated in the 4f Wyckoff site with fractional coordinates (2/3, 1/3, 0.13336). The Ti atoms are located on the 2a Wyckoff position with fractional coordinates (0, 0, 0). The Al atoms are positioned in the 2b Wyckoff site with fractional coordinates (0, 0, 1/4). The atomic positions for C is in 4f Wyckoff position with fractional coordinates (1/3, 2/3, 0.0687). The calculated values of structural properties of $Mo_2TiAlC_2$ are presented in Table 1 along with the available experimental results. As can be seen from Table 1, the theoretical results are very close to the experimental values. This ensures the reliability of the present DFT-based first-principles calculations.

Table 1. Lattice constants $a$ and $c$, internal parameter z, hexagonal ratio c/a, unit cell volume V for $Mo_2TiAlC_2$.

| Properties | $a$ (Å) | $c$ (Å) | $Z_{Mo}$ | $Z_C$ | $c/a$ | $V$ (Å$^3$) |
|---|---|---|---|---|---|---|
| Expt. [1] | 2.997 | 18.661 | 0.1334 | 0.0687 | 6.2264 | 145.16 |
| Present Calc. | 2.998 | 18.752 | 0.1332 | 0.0686 | 6.2548 | 145.92 |

### 3.2 *Elastic properties*

Table 2 lists our results for elastic properties at zero pressure. No experimental data on elastic properties for $Mo_2TiAlC_2$ are available. It is found that the calculated elastic constants are all positive and satisfy the well-known Born criteria [28] for stability of hexagonal system: $C_{11} > 0$, $(C_{11}-C_{12}) > 0$, $C_{44} > 0$ and $(C_{11}+C_{12}) C_{33} > 2C_{13}^2$. The small values of $C_{12}$ and $C_{13}$ also imply that $Mo_2TiAlC_2$ should be brittle in nature [29].The theoretical polycrystalline elastic moduli (bulk moduli,$B$, compressibility,$K$, shearmoduli,$G$, Young'smoduli,$Y$, Pough ratio and the Poissonratio $v$) are given in Table 2. $Y$ and $v$ are computed using the relationships: Y=9BG/(3B+G) and $v$ =(3B-Y)/6B [30]

Though the bulk modulus $B$ is not the measure of the hardness of the materials but according to the CASTEP code bulk modulus for hard materials is normally greater than 200 GPa. Therefore one can say that $Mo_2TiAlC_2$ is hard materials with $B = 222$ GPa. The Young's modulus $Y$, denoting a measure of stiffness, has the value of 331 GPa indicating stiff nature of the material.

The Pugh's criterion [31] and the Frantsevich's rule [32] also support this classification. According to Pugh's criterion, a material should be brittle if its Pugh's ratio G/B > 0.5, otherwise it should be ductile. The Frantsevich's rule has suggested the Poisson's ratio $\mu \sim 0.33$ as the critical value that separates the brittle and the ductile behavior. If the Poisson ratio $\mu$ is less than 0.33, the mechanical property of the material is dominated by brittleness, if larger than 0.33, the mechanical property of the material mainly shows ductility. From Table 2, it is evident that the compound $Mo_2TiAlC_2$ is brittle in nature that shows general trend of MAX phases. The relatively low value of the Poisson's ratio for $Mo_2TiAlC_2$ is indicative of its high degree of directional covalent bonding.

Table 2. Calculated elastic constants ($C_{ij}$, in GPa), Bulk modulus $B$ (GPa), Shear modulus $G$ (GPa), Young's modulus $Y$ (GPa), Shear modulus to Bulk modulus ratio $G/B$, Poisson's ratio $v$, Elastic anisotropic factor $A$ and linear compressibility ratio $k_c/k_a$ for $Mo_2TiAlC_2$ for $Mo_2TiAlC_2$.

| Compound | $C_{11}$ | $C_{12}$ | $C_{13}$ | $C_{33}$ | $C_{44}$ | $C_{66}$ | $B$ | $G$ | $Y$ | $G/B$ | $v$ | $A$ | $k_c/k_a$ |
|---|---|---|---|---|---|---|---|---|---|---|---|---|---|
| $Mo_2TiAlC_2$ | 391 | 153 | 138 | 366 | 154 | 119 | 222 | 133 | 331 | 0.60 | 0.251 | 1.30 | 1.10 |

The difference between $C_{11}$ and $C_{33}$ suggests that $Mo_2TiAlC_2$ possesses anisotropy in elastic properties. Elastic anisotropy of a crystal reflects different characteristics of bonding in different directions. Essentially almost all the known crystal are elastically anisotropic, and a proper description of such an anisotropic behavior has, therefore, an important implication in engineering science and crystal physics since it correlates with the possibility of appearance of microcracks inside the crystal. The shear anisotropy factor for the {100} shear planes between the [011] and [010] directions is given by $A = 4C_{44}/(C_{11}+C_{33}-2C_{13})$. For an isotropic crystal $A = 1$, while deviation from ~~(any value smaller or greater than)~~ unity is a measure of the degree of elastic anisotropy possessed by the crystal. The calculated shear anisotropic factor shown in Table 2 deviates from unity, which implies that the in-plane and out-of-plane inter-atomic interactions differ somewhat. Another anisotropic factor is defined by the ratio between the linear compressibility coefficients along the $c$- and $a$-axis for the hexagonal crystal: $k_c/k_a = (C_{11} + C_{12} - 2C_{13})/(C_{33} - C_{13})$. This has also been calculated. The calculated value of 1.10 reveals that the compressibility along the $c$-axis is larger than that along the $a$-axis for this compound.

### 3.3 Electronic properties

The band structure of $Mo_2TiAlC_2$ with optimized lattice parameters is presented in Fig. 1 along the high symmetry directions in the first Brillouin zone. The Fermi level is taken at zero of the energy scale. From figure it is found that the valence band and conduction band overlapped considerably which exhibits the conducting nature of the system. The occupied valence bands spread widely from -12.8 eV to the Fermi level $E_F$. The highly non-dispersive nature of the band structure along $\Gamma$-$A$ direction shows that electronic transport along c-direction is not that different from that in the plane.

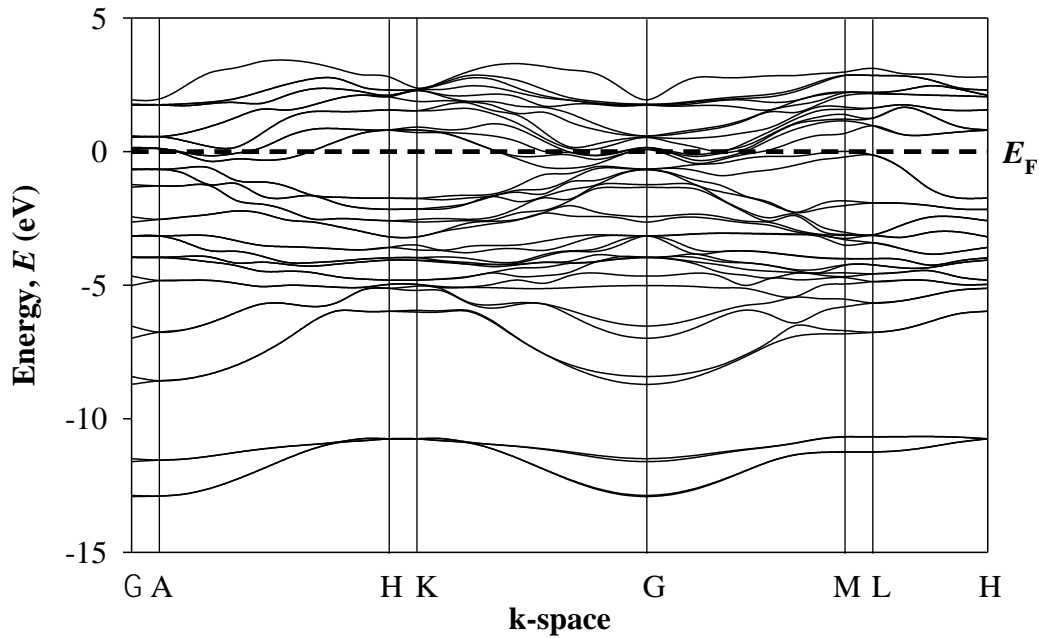

Fig.1. Band structure of $Mo_2TiAlC_2$

To further elucidate the nature of conductivity we have calculated the total and partial density of states (DOS) and displayed in Fig. 2. Normally the conductivity is proportional to the density of states. The high value of DOS (~ 6 states/eV) indicating also the good conducting nature of this compound. From the partiol DOS we see that the energy bands around the Fermi level are mainly from the Mo 4$d$ states. This indicates that the Mo 4$d$ states dominate the conductivity of Mo$_2$TiAlC$_2$ but the contribution from Ti 3d states is also considerable. From -15 to -7 eV energy range the contribution to the DOS mainly comes from the C 2$s$ states, while the little contribution from the Ti 3$d$ and Mo 4d states are also noticeable The energy bands between -7 to 0 eV are dominated by hybridized Mo 4d/4p, Ti 3$d$, Al 3$s$/3$p$ and C 2$p$ states. It is found that Mo-Ti bond is stronger than Ti-Al and Al- C in the compound Mo$_2$TiAlC$_2$.

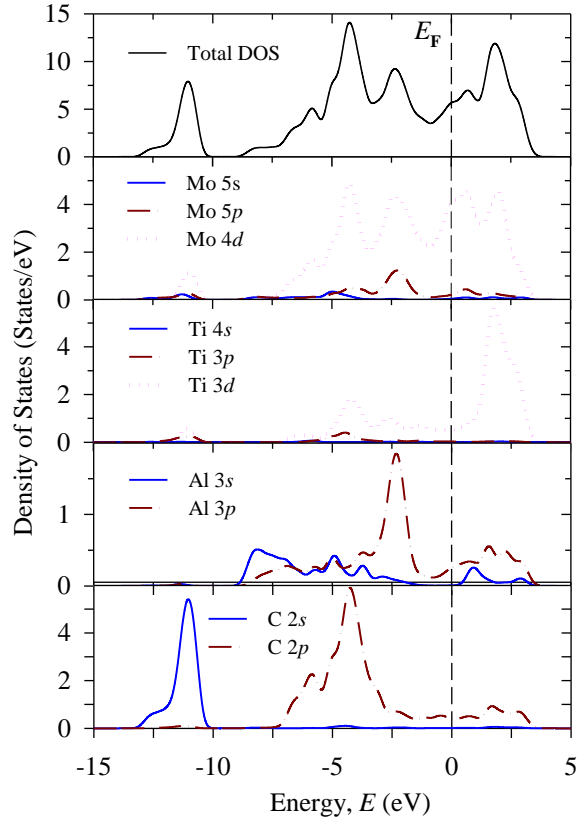

Fig.2.Total and partial density of states (DOS) of Mo$_2$TiAlC$_2$

### 3.4 *Optical properties:*

When an electromagnetic radiation is incident on the materials, different materials response in different nature. Optical properties of a material mean a material's response to electromagnetic radiation and, in particular, to visible light. The optical properties of Mo$_2$TiAlC$_2$ can be obtained from the complex dielectric function; $\varepsilon(\omega) = \varepsilon_1(\omega) + i\varepsilon_2(\omega)$ which is one of the main optical characteristics of solids. The expression for the imaginary part $\varepsilon_2(\omega)$ can be found elsewhere [33-35] which is used in CASTEP [20] to obtain its numerically by a direct evaluation of the matrix elements between the occupied and unoccupied electronic states. The real part $\varepsilon_1(\omega)$ of dielectric function can be derived from the imaginary part $\varepsilon_2(\omega)$ by the Kramers-Kronig relations. The other optical properties, such as refractive index, absorption spectrum, loss-function, reflectivity, and conductivity (real part) are derived from $\varepsilon_{1}(\omega)$ and $\varepsilon_{2}(\omega)$ [20].

The optical properties are closely related to the electronic band structure of the materials and used to explain the electronic response of the materials to the electromagnetic radiation. The optical constants of $Mo_2TiAlC_2$, e.g., the real part and imaginary part of dielectric function $\varepsilon_1$ and $\varepsilon_2$, refractive index, extinction co-efficient, absorption, energy-loss function, reflectivity and photoconductivity are shown in Figs. 3. We used a 0.5 eV Gaussian smearing for all calculations. This smears out the Fermi level, so that k-points will be more effective on the Fermi surface.

When light of sufficient energy shines onto a material, it induces transitions of electrons from occupied states below the Fermi energy to unoccupied states above the Fermi energy. In metal and metal-like systems the intraband contribution to the optical properties affects mainly the low energy infrared part of the spectra. In order to calculate the dielectric function a semiempirical Drude term is employed. A Drude term with unscreened plasma frequency 3 eV and damping 0.05 eV has been used. The peak of the imaginary part of the dielectric function is related to the electron excitation. It is observed in the imaginary part $\varepsilon_2$ of the dielectric function that the peak around 1 eV is due to transitions within the Mo 4d bands. The large negative value of $\varepsilon_1$ indicates that the $Mo_2TiAlC_2$ crystal has a Drude-like behavior.

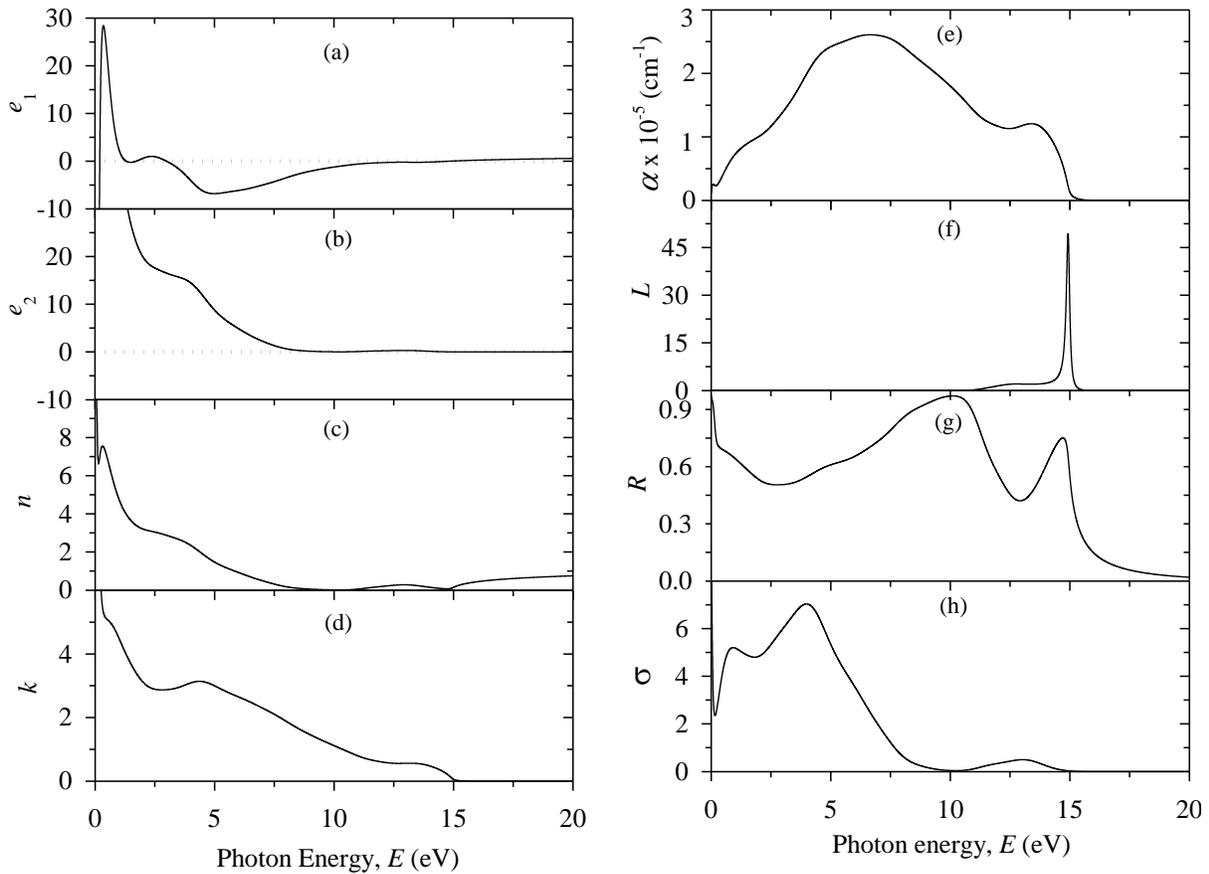

Fig. 3. Optical constant of $Mo_2TiAlC_2$: (a) Real part $\varepsilon_1$ of dielectric function, (b) Imaginary part $\varepsilon_2$ of the dielectric function, (c) Refractive index, n, (d) Extinction co-efficient, k, (e) Absorption spectrum, (f) Loss function, (g) Reflectivity, and (h) Real part of Conductivity.

The refractive index is another technically important parameter for optical materials for its use in optical devices such as photonic crystals, waveguides etc. The refractive index $n$ and the extinction coefficient $k$ are displayed in Figures 3(c) and 3(d), respectively. The static refractive index $n(0)$ is found to have the value of ~ 9.

Figure 3e shows the absorption coefficient spectra of the compound under consideration which reveal the metallic nature of the compound since the spectrum starts with non-zero value and rises to maximum value at around 7eV. The absorption spectrum falls sharply at the peak of loss function.

The loss function $L(\omega)$, shown in Fig. 3f, which unfolds the energy loss of a fast electron traversing in the material. Its peak is defined as the bulk plasma frequency $\omega_P$, which occurs at $\varepsilon_2 < 1$ and $\varepsilon_1 = 0$. In the energy-loss spectrum, we see that the effective plasma frequency $\omega_P$ of the compound under consideration is equal to ~ 15 eV. When the incident photon frequency is higher than $\omega_P$, the material becomes transparent. The reflectivity spectra as a function of photon energy are shown in Fig. 3g. It is found that the reflectivity of the compound starts with a value of ~ 0.75, decreases and then rises again to reach maximum value of ~ 0.95 between ~ 2.3 and 10 eV. The large reflectivity in very low energy range indicates the characteristics of high conductance in the low energy region. Moreover, the peak of loss function corresponds to the trailing edges in the reflection spectra. Since the material have no band gap as evident from band structures, the photoconductivity starts with zero photon energy for each of the phase as shown in Figure 3h. The spectra have several maxima and minima within the energy range studied.

## 4. Conclusions

First-principles calculations based on DFT have been used to investigate the mechanical, electronic and optical properties of the new compound, $Mo_2TiAlC_2$, for the first time. The equilibrium crystal structures of this compound are in good agreement with experimental values. The mechanical stability has been confirmed by elastic constants. The analysis of elastic constants also reveals that the compound is hard, stiff and brittle in nature. The compressibility along the $c$-axis is larger than that along the $a$-axis for $Mo_2TiAlC_2$. The electronic band and DOS shows metallic behavior of this compound. The dielectric function, refractive index, absorption spectra, energy loss function, reflectivity and conductivity are determined and analyzed in detail. The reflectivity of this compounds is high (~95%) in the IR-Visible-UV region which indicates that the material is a potential coating material to avoid solar heating. The ability of the reflection of this compound is stronger compared to similar compounds. The study should provide incentives for further experimental investigation which would have the way for practical application.